
\hcorrection{36pt}
\TagsOnRight
\magnification=\magstep1
\loadbold

\define\dd#1{\partial_#1}
\define\du#1{\partial^#1}
\define\ndd#1#2{\eta_{#1#2}}
\define\nuu#1#2{\eta^{#1#2}}
\define\lc#1#2#3#4{\varepsilon_{#1#2#3#4}}
\define\drt#1#2{\frac{\partial#1}{\partial#2}}
\define\gdd#1#2{g_{#1#2}}
\define\guu#1#2{g^{#1#2}}
\define\ru{R{\cdot}u}
\define\rup#1{(R{\cdot}u)^#1}
\define\rs{R{\cdot}s}
\define\rsp#1{(R{\cdot}s)^#1}
\define\pos{\boldkey r}
\define\aalign{\spreadlines{1\jot}\allowdisplaybreaks\align}
\documentstyle{amsppt}
\topmatter
\leftheadtext{Jaegu Kim}
\rightheadtext{Gravitational Field of a Moving Spinning Point Particle}
\title
Gravitational Field of \\
a Moving Spinning Point Particle
\endtitle
\author
Jaegu Kim\footnote"*"{jaegukim\@cc.kangwon.ac.kr}
\endauthor
\affil
Department of Physics,
Kangwon National University,
Chunchon 200--701 Korea
\endaffil
\abstract
The gravitational and electromagnetic fields of a moving charged spinning
point particle are obtained in the Lorentz covariant form by transforming the
Kerr--Newman solution in Boyer--Lindquist coordinates to the one in the
coordinate system which resembles the isotropic coordinates and then
covariantizing it. It is shown that the general relativistic proper time at
the location of the particle is the same as the special relativistic one and
the gravitational and electromagnetic self forces vanish.
\flushpar
\endabstract
\endtopmatter

\voffset=-0.25in

\document
\bigskip\flushpar
{\bf 1. Introduction}
\medskip\par
In the previous paper [1] we developed a Lorentz covariant method of solving
the Einstein equations and obtained the gravitational field of a moving point
particle for both uncharged and charged cases.  It is shown that the general
relativistic proper time at the location of the particle is the same as the
special relativistic one and the gravitational and electromagnetic self forces
vanish. Thus, the solution consistently describes the gravitational and
electromagnetic fields of a charged point particle. We have seen that the
isotropic coordinates played a crucial role in the description of the source
as a point particle.
\par
In this paper we will consider the gravitational and electromagnetic fields of
a moving charged spinning point particle. For a spinless point particle we
can express the metric tensor simply in terms of a Lorentz invariant quantity
$\ru$ and two Lorentz covariant tensors $\ndd\mu\nu$ and $u_\mu u_\nu$. But we
cannot expect this kind of simplicity for a spinning point particle.
For a spinning point particle moving with constant velocity,
we can introduce one more Lorentz four vector $s^\mu$ which is a four vector
generalization of spin angular momentum per unit mass. Then the Lorentz
covariant quantities which can be used to make an ansatz for the metric
tensor are $\ndd\mu\nu$, $\lc\mu\nu\sigma\lambda$, $R^\mu$, $u^\mu$, and
$s^\mu$, from which one can construct two Lorentz invariant quantities and
ten symmetric Lorentz covariant tensors of rank two. Thus, the most general
expression for the metric tensor is so complicated that it is almost impossible
to calculate the Ricci tensor and solve the Einstein--Maxwell equations
directly. To avoid this complication we reverse the procedure described in [1]
instead of solving the Einstein--Maxwell equations directly. We start with the
Kerr--Newman solution in Boyer--Lindquist coordinates and transform the
coordinate system to the one which resembles the isotropic coordinates.
Then we let the particle move with constant velocity and obtain the metric
tensor and the four vector potential by covariantizing them.
Using computer, we confirm that the metric tensor and the four vector potential
satisfy the Einstein--Maxwell equations by substituting them into the
field equations. Finally, we check the consistency of the solution and its
implications. We will use the same notational conventions given in [1].
\bigskip\flushpar
{\bf 2. Metric tensor and four vector potential}
\medskip\par
The line element and the potential one form for a source of mass $m$, spin
angular momentum $J=ms$ and charge $q$ is given by the Kerr--Newman solution
in Boyer--Lindquist coordinates [2] as
$$ \aalign
   ds^2&=-\left[1 - \dfrac{(2m\rho - q^2)}\Sigma\right]dt^2 -
   \frac{2(2m\rho - q^2)s\sin^2\theta}\Sigma\,dt\,d\phi +
   \frac\Sigma\Delta\, d\rho^2\\
   &\qquad+\Sigma\,d\theta^2+\left[\rho^2 + s^2 +
   \frac{(2m\rho - q^2)s^2\sin^2\theta}\Sigma\right]\sin^2\theta\,d\phi^2,
   \tag2.1a \\
   A_\mu\,&dx^\mu = -\frac{q\rho}\Sigma(dt - s\sin^2\theta\,d\phi), \tag2.1b
   \endalign $$
where the source is spinning in the $\phi$ direction and
$$ \align
   \Delta&=\rho^2-2m\rho+s^2+q^2, \tag2.2a\\
   \Sigma&=\rho^2+s^2\cos^2\theta. \tag2.2b
   \endalign $$
\par
Recalling that the isotropic coordinates played a crucial role in the
description of the source as a point particle [1], we transform the coordinate
system by introducing a new radius $r$ such that the transformed metric
reduces to the Schwarzschild metric in isotropic coordinates when $s = q =0$
$$ \rho=f(r). \tag2.3 $$
The condition for isotropy is
$$  {f'}^2=\frac\Delta{r^2}, \tag2.4 $$
and its solution is found to be
$$ \rho=\biggl(1+\frac mr +\frac{m^2 - s^2 -q^2}{4r^2}\biggr)r. \tag2.5 $$
Using this transformation rule, we have
\vfil\eject\flushpar
$$ \aalign
   ds^2&=-(1-2\Phi)dt^2 -4s\Phi\sin^2\theta\,dt\,d\phi \\
   &\quad+\biggl[\biggl(1+\frac mr+\frac{m^2-s^2-q^2}{4r^2}\biggr)^2 +
   \frac{s^2\cos^2\theta}{r^2}\biggr](dr^2+r^2 d\theta^2+r^2\sin^2\theta\,
   d\phi^2)\\
   &\quad + (1+2\Phi)s^2\sin^4\theta\,d\phi^2, \tag2.6a\\
   A_\mu\,&dx^\mu =\frac{\dfrac qr\biggl(1+ \dfrac mr + \dfrac{m^2 - s^2 - q^2}
   {4r^2}\biggr)}{\biggl(1 + \dfrac mr + \dfrac{m^2 - s^2 - q^2}{4r^2}\biggr)^2
   + \dfrac{s^2\cos^2\theta}{r^2}}(-dt + s\sin^2\theta\,d\phi), \tag2.6b
   \endalign $$
where
$$ \Phi=\frac{\dfrac mr\left(1 + \dfrac mr + \dfrac{m^2 - s^2 - q^2}{4r^2}
   \right) - \dfrac{q^2}{2r^2}}{\left(1 + \dfrac mr + \dfrac{m^2 - s^2 - q^2}
   {4r^2}\right)^2 + \dfrac{s^2\cos^2\theta}{r^2}}. \tag2.7 $$
{}From now on we will refer to the coordinate system in (2.6) as the isotropic
coordinate system. Next we transform the spherical coordinates $(r,\theta,
\phi)$ to the Cartesian coordinates $(x^1,x^2,x^3)$
$$ \aalign
   x^1&= r\sin\theta\cos\phi,\tag2.8a\\
   x^2&= r\sin\theta\sin\phi,\tag2.8b\\
   x^3&= r\cos\theta.\tag2.8c
   \endalign $$
Then the quantities in (2.6) can be written as
$$ \aalign
   &s\cos\theta=\frac{\boldkey s{\cdot}\pos}r, \tag2.9a\\
   &dr^2+r^2 d\theta^2+r^2\sin^2\theta\,d\phi^2=\ndd i j\,dx^i dx^j, \tag2.9b\\
   &s r^2\sin^2\theta\,d\phi=\varepsilon_{ijk}s^i x^j dx^k. \tag2.9c
   \endalign $$
\par
Now let the source move with constant four velocity $u^\mu$. Then we can obtain
new line element and potential one form by covariantizing the quantities in
(2.6) as follows:
$$ \aalign
   &r\rightarrow -\ru, \tag2.10a\\
   &s\cos\theta\rightarrow -\frac{\rs}{\ru}, \tag2.10b \\
   &dt\rightarrow -u_\mu dx^\mu, \tag2.10c\\
   &dr^2+r^2 d\theta^2+r^2\sin^2\theta\,d\phi^2\rightarrow
   (\ndd\mu\nu + u_\mu u_\nu)dx^\mu dx^\nu,\tag2.10d\\
   &s r^2\sin^2\theta\,d\phi\rightarrow \lc\mu\nu\sigma\tau
   u^\mu s^\nu R^\sigma dx^\tau, \tag2.10e
   \endalign $$
where $s^\mu$ is four spin angular momentum vector per unit mass
$$ s^\mu=\bigl(\boldkey s{\cdot}\boldkey u,\,\boldkey s +
   \frac{(\boldkey s{\cdot}\boldkey u)\boldkey u}{\gamma+1}\bigr). \tag2.11 $$
Here $\boldkey s$ is the spin angular momentum vector per unit mass in the rest
frame of the source. From the new line element and potential one form we
construct the metric tensor and the four vector potential in the Lorentz
covariant form
$$ \aalign
   \gdd\mu\nu &= e^A[\ndd\mu\nu + (1-e^B)u_\mu u_\nu] -
   \frac{(1-e^{A+B})}{\rup2}(u_\mu k_\nu + u_\nu k_\mu) +
   \frac{(2-e^{A+B})}{\rup4}k_\mu k_\nu, \tag2.12a\\
   A_\mu &= -\frac{q\,e^{-A}}\ru\biggl[1 - \dfrac m\ru + \dfrac{m^2- s^2- q^2}
   {4\rup2}\biggr]\biggl[u_\mu - \frac{k_\mu}{\rup2}\biggr],\tag2.12b
   \endalign $$
where
$$ \aalign
   e^A&=\biggl[1 - \frac m\ru + \frac{m^2 - s^2 - q^2}{4\rup2}\biggr]^2 +
   \frac{\rsp2}{\rup4}, \tag2.13a\\
   e^{A+B}&=\frac{\biggl[1 - \dfrac{m^2 - s^2 - q^2}{4\rup2}\biggr]^2 +
   \dfrac{\rsp2}{\rup4} - \dfrac{s^2}{\rup2}}{\biggl[1 - \dfrac m\ru +
   \dfrac{m^2 - s^2 - q^2}{4\rup2}\biggr]^2 + \dfrac{\rsp2}{\rup4}},\tag2.13b\\
   k_\mu &= \lc\mu\alpha\beta\gamma u^\alpha s^\beta R^\gamma.
   \tag2.13c
   \endalign $$
Note that the metric tensor and the four vector potential (2.12) reduce to
those of spinless case [1] when $s$ is zero. Since the vector $k_\mu$
satisfies
$$ \aalign
   k_\mu k_\nu &=[s^2\rup2-\rsp2]\ndd\mu\nu- s^2 R_\mu R_\nu -
   s^2(\ru)(R_\mu u_\nu +  R_\nu u_\mu)\\
   &\quad+(\rs)(R_\mu s_\nu +R_\nu s_\mu)  - \rsp2 u_\mu u_\nu \\
   &\quad + (\ru)(\rs) (u_\mu s_\nu +u_\nu s_\mu) - \rup2 s_\mu s_\nu, \tag2.14
   \endalign $$
we could replace the product $k_\mu k_\nu$ in (2.12a) by the expression on the
right hand side of (2.14), but we will retain the expression (2.12a) for
simplicity. Calculating the determinant of the metric tensor and the
contravariant metric tensor, we have
$$ \aalign
   g&= -e^{2A+C}, \tag2.15a\\
   g^{\mu\nu}&=e^{-A}\eta^{\mu\nu} + e^{-A}[(1-e^{-B}) + (1-e^{A+B})^2
   (e^{-B}-e^{2A-C})]u^\mu u^\nu\\
   &\quad + \frac{e^{-C}(1-e^{A+B})}{\rup2}(u^\mu k^\nu + u^\nu k^\mu) -
   \frac{e^{-A-C}}{\rup4}k^\mu k^\nu, \tag2.15b
   \endalign $$
where
\vfil\eject\flushpar
$$ \aalign
   e^C &= e^{2A+B}+\dfrac{s^2}{\rup2}-\dfrac{\rsp2}{\rup4} \\
   &= \biggl[1-\frac{m^2-s^2-q^2}{4\rup2}\biggr]^2. \tag2.16
   \endalign $$
\bigskip\flushpar
{\bf 3. Field equations}
\medskip
Instead of solving the Einstein--Maxwell directly we have constructed the
metric tensor and the four vector potential by transforming the Kerr--Newman
solution in Boyer--Lindquist coordinates to the one in the isotropic
coordinates and then covariantizing it. If (2.12) satisfies the
Einstein--Maxwell equations, it immediately implies that (2.1) also satisfies
the Einstein--Maxwell equations. But the converse is not necessarily true.
Hence we have to show explicitly that (2.12) satisfies the Einstein--Maxwell
equations. However, due to the complexity of the metric tensor (2.12) the
Christoffel symbol turns out to be extremely complicated.
Moreover, since the Ricci tensor has terms of products of two
Christoffel symbols, it is almost impossible to calculate the Ricci tensor
by hand. Nevertheless, we confirmed that (2.12) satisfied the Einstein--Maxwell
equations by using computer with the symbol manipulating language
Mathematica [3].
\bigskip\flushpar
{\bf 4. Discussion}
\medskip
We have obtained the metric tensor and the four vector potential of a
moving charged spinning point particle in the Lorentz covariant form. To check
consistency of the solution with the equations of motion, we consider the
gravitational and electromagnetic self forces and find
$$ \aalign
   F_{\text{grav}}^\lambda&=-m[\Gamma^\lambda_{\mu\nu}]_{\vphantom{q}_{R=0}}
   u^\mu u^\nu =0, \tag4.1a\\
   F_{\text{em}}^\lambda&=q[F^\lambda{}_\sigma u^\sigma]_{\vphantom{q}_{R=0}}
   =0. \tag4.1b
   \endalign $$
Hence the gravitational and electromagnetic self forces vanish. Next we
consider the general relativistic proper time interval at the location of the
particle. Except for the case $m^2=s^2+q^2$, we have
$$ d\tau =\frac1\gamma[-\gdd\mu\nu u^\mu u^\nu]^{1/2}_{\vphantom{q}_{R=0}}
   dt = \frac1\gamma dt, \tag4.2 $$
i\.e\., it is the same as the special relativistic proper time interval. For
the case $m^2=s^2+q^2$, it is indeterminate. In this case one can
choose the special relativistic proper time as the Affine parameter of the
trajectory of the particle. Therefore, the solution (2.12) is consistent with
the equations of motion and properly describes the source as a spinning
point particle.
\par
As we have seen, the most natural coordinate system for a charged spinning
point particle is the isotropic coordinates rather than the Boyer--Lindquist
coordinates. The topology of base manifold is trivial in the isotropic
coordinates. We now turn to the coordinate transformation (2.5) and consider
the global topology in Boyer--Lindquist coordinates induced by it. We can
classify the transformation (2.6) into three cases according to the value of
$m^2 - s^2 - q^2$.
\par
First, consider the case with $m^2>s^2 + q^2$, which may hold for an
astrophysical object after gravitational collapse. Due to the nonlinearity of
the transformation (2.6) the region between $\rho=m+(m^2-s^2-q^2)^{1/2}$ and
$\rho=\infty$ is mapped twice, while the region inside
$\rho=m+(m^2-s^2-q^2)^{1/2}$ is not mapped by any value of $r$. Thus the
transformation (2.6) induces a nontrivial topology in Boyer--Lindquist
coordinates. To visualize the global topology of spacetime in Boyer--Lindquist
coordinates we consider embedding of a spacelike hypersurface of constant
time $t=0$, with one degree of rotational freedom suppressed ($\theta=\pi/2$).
In Boyer--Lindquist coordinates this hypersurface corresponds to two distinct
asymptotically flat hypersurfaces which are connected by the Einstein--Rosen
bridge at $\rho=m+(m^2-s^2-q^2)^{1/2}$, which is the same as the topology
of the Schwarzschild geometry in Schwarzschild coordinates [1].
The location of the particle $r=0$ is mapped to $\rho=\infty$
where spacetime is flat.
\par
Second, consider the case with $m^2=s^2+q^2$. Note that the minimum value of
$\rho$ in (2.5) is $m$. Hence the region inside $\rho=m$ in Boyer--Lindquist
coordinates should be excised. The location of the particle $r=0$ is mapped
to $\rho=m$.
\par
Last, consider the case with $m^2<s^2+q^2$, which holds for
elementary particles with spin or charge.  For instance estimating the
numerical values for these quantities for an electron, we have
$m^2=4.6{\times}10^{-115}\,\text{m}^2$,
$s^2=1.1{\times}10^{-25}\,\text{m}^2$,
and $q^2=1.9{\times}10^{-72}\,\text{m}^2$. Thus, $s$ plays dominant role
in the behavior of the gravitational field of the electron at short distance.
We note that the metric tensor and the four vector potential has a ring
singularity at $r=[(s^2+q^2)^{1/2}-m]/2$ and $\theta=\pi/2$.
In this case the lower limit of $\rho$ should be extended
to $-\infty$ instead of 0 and the location of the particle $r=0$ is mapped to
$\rho=-\infty$ where spacetime is flat.
\par
To understand the short distance behavior of the gravitational and
electromagnetic forces we consider a test particle of mass $m_0$ and charge
$q_0$ near the static spinning point source described by (2.6). Since the
gravitational and electromagnetic forces in general have a very complicated
dependence on distance, we will consider a static test particle on the
equatorial plane ($\theta=\pi/2$) for simplicity. In this case both forces
on the test particle point in the radial direction and turn out to be
$$ \aalign
   F_{\text{g}} &= -m_0 \Gamma^r_{tt}\\
   &= -\frac{m_0m}{r^2}\frac{\biggl(\mathstrut 1 - \dfrac{m^2-s^2-q^2}{4r^2}
   \biggr)\biggl(\mathstrut 1 + \dfrac{m^2-q^2}{mr} + \dfrac{m^2-s^2-q^2}
   {4r^2}\biggr)}{\biggl(1 + \dfrac mr + \dfrac{m^2-s^2-q^2}{4r^2}\biggr)^5},
   \tag4.3a\\
   F_{\text{e}} &= q_0 F^r{}_t\\
   &= \frac{q_0q}{r^2}\frac{\biggl(\mathstrut 1 - \dfrac{m^2-s^2-q^2}{4r^2}
   \biggr)}{\biggl(1 + \dfrac mr + \dfrac{m^2-s^2-q^2}{4r^2}\biggr)^4}.
   \tag4.3b
   \endalign $$
We note that $F_{\text{g}}$ and $F_{\text{e}}$ are inversely proportional to
$r^2$ near $r=\infty$. Now let the test particle move from $r=\infty$ to $r=0$
and consider the forces on it for three different cases separately.
\par
First, consider the case with $m^2>s^2+q^2$. As the test particle approaches
$r=(m^2-s^2-q^2)^{1/2}/2$, both $F_{\text{g}}$ and $F_{\text{e}}$ approach
zero and reverse their directions when it passes $r=(m^2-s^2-q^2)^{1/2}/2$.
Thus, at short distance the gravitational force becomes repulsive and
the electrostatic force becomes attractive for the same type of charges and
repulsive for opposite type of charges. Both $F_{\text{g}}$ and
$F_{\text{e}}$ are proportional to $r^4$ near $r=0$.
\par
Second, consider the case with $m^2=s^2+q^2$. In this case $F_{\text{g}}$ and
$F_{\text{e}}$ never change their directions and are proportional to $r^2$ near
$r=0$.
\par
Last, consider the case with $m^2<s^2+q^2$. In this case both $F_{\text{g}}$
and $F_{\text{e}}$ are infinite at $r=[(s^2+q^2)^{1/2}-m]/2$ due to the ring
singularity and $F_{\text{e}}$ never changes its direction. For $q=0$
$F_{\text{g}}$ never changes its direction, but for $q\ne 0$ $F_{\text{g}}$
changes its direction twice at $r=[(s^2-q^2+q^4/m^2)^{1/2}+q^2/m -m]/2$ and at
\hbox{$r=[(s^2+q^2)^{1/2}-m]/2$}. Thus, for $q\ne 0$ the gravitational force
becomes repulsive in the region between $r=[(s^2+q^2)^{1/2}-m]/2$ and
$r=[(s^2-q^2+q^4/m^2)^{1/2}+q^2/m -m]/2$. Both $F_{\text{g}}$ and
$F_{\text{e}}$ are proportional to $r^4$ near $r=0$.
\par
We have seen that both the gravitational and electrostatic forces on the static
test particle on the equatorial plane approach zero as the test particle
approaches $r=0$. Although the behavior of the gravitational and
electromagnetic forces on the nonstatic test particle off the equatorial plane
is very complicated, we also have confirmed that they both approach zero as
the test particle approaches $r=0$. Therefore the gravitational and
electromagnetic forces are asymptotically free at short distance.
This sheds a new light on the theory of quantum gravity, since quantum
theory is less divergent than classical theory.
\bigskip\flushpar
{\bf Acknowledgement}
\medskip
I would like to thank Professor Robert Finkelstein for reading the manuscript
and comments and Professors Young Jik Ahn and Taejin Lee for helpful
discussions.
\bigskip\flushpar
{\bf References}
\medskip
\item{[1]} Kim J 1993 ``{\it Gravitational Field of a Moving Point Particle}"
\hbox{gr--qc/9311027} (to be published)
\item{[2]} Boyer R H and Lindquist R W 1967 {\it J\. Math\. Phys\.} {\bf 8}
265
\item{[3]} Wolfram S 1991 ``{\it Mathematica\/}: {\it A System for Doing
Mathematics by Computer}" 2nd Ed\. (Addison--Wesley)
\enddocument